\documentstyle[12pt]{article}
\begin{document}
\title{
Path Integral for Relativistic 
Dionium System}
\author{De-Hone Lin \thanks{e-mail:d793314@phys.nthu.edu.tw}
\vspace{0.5cm} \\ Department of Physics,
National Tsing Hua University \\
Hsinchu 30043, Taiwan, Republic of China \\}
\maketitle
\setlength{\baselineskip}{1cm} \centerline{\bf Abstract}
{The path integral for relativistic 
spinless dionium atom 
is solved, and the energy spectra 
are extracted from the resulting amplitude.}
\thispagestyle{empty} \newpage
\renewcommand{\thesection}{\arabic{section}}

\tolerance=10000

\section{INTRODUCTION}

~~~~~~~~~Although 
the idea of magnetic monopoles probably was discussed in classic
physics early in the history of electricity and magnetism, modern discussions
of this concept date back only to 1931 by Dirac \cite{1}. He pointed out
that magnetic monopoles in quantum mechanics exhibit some extra and subtle
features. In particular, with the existence of a magnetic monopole of
strength $g$, electric charges and magnetic charges must necessarily be
quantized, in quantum mechanics \cite{1,yang}.
Nowadays, we know that the quantization
condition $2eg/\hbar c=$integer exists even at the classical level.
This is explained 
by using the extra monopole gauge invariance \cite{2}. 
This invariance expresses the physical
irrelevance of the shape of the Dirac strings attached to the monopoles. 
Employing the invariance, the quantization condition
can be proved for any fixed particle orbits, i.e. without invoking
fluctuating orbits which would correspond to the standard derivation using
Schr\"odinger wave functions. 

In modern elementary 
particle and cosmology theories, magnetic monopole plays
an important role. All the grand unified theories predict the existence
of monopoles, so do all of the cosmology theories \cite{3}.
Therefore the problems relating to the monopoles are always interesting to the
physicists. In this paper, we calculate the path integral (PI) of the relativistic
dionium atom, i.e., a system of two relativistic particles with both electric 
and magnetic charges $(e_1, g_1)$ and $(e_2, g_2)$. 

In the past 15 years, considerable progress has been made in solving path
integral of potential problems \cite{4,5}. However, only a few relativistic
problems have been discussed using 
PI \cite{6,7,8,8.1,8.2,11,12,12.1}. In this paper, we
solve, following Kleinert's method \cite{2.1}, the relativistic spinless
3-dimensional dionium system by path integral. The energy spectra are
extracted from the resulting amplitude.

\section{THE RELATIVISTIC PATH INTEGRAL}

~~~~~~~~~Adding a 
vector potential ${\bf {A}({x}})$ to Kleinert's relativistic path
integral for a particle in a potential $V({\bf {x}})$ \cite{6}, we find
that the
expression of the fixed-energy amplitude is \cite{8.1}
\begin{equation}
\label{2.1}G({\bf {x}}_b,{\bf {x}}_a;E)=\frac{i\hbar }{2Mc}\int_0^\infty
dL\int D\rho \Phi \left[ \rho \right] \int D^Dxe^{-A_E/\hbar } 
\end{equation}
with the action 
\begin{equation}
\label{2.2}A_E=\int_{\tau _a}^{\tau _b}d\tau \left[ \frac M{2\rho \left(
\tau \right) }x^{\prime ^2}\left( \tau \right) -i{A\cdot x^{\prime }(\tau }%
)-\rho (\tau )\frac{\left( E-V\right) ^2}{2Mc^2}+\rho \left( \tau \right) 
\frac{Mc^2}2\right] . 
\end{equation}
For the dionium system under consideration, the potential is 
\begin{equation}
\label{2.3}-e^2/r, 
\end{equation}
and the vector potential 
\begin{equation}
\label{2.4}{\bf A(x})=\hbar q\frac{(x_1{\bf \hat x}_2-x_2{\bf \hat x}_1)x_3}{%
r{\bf x}_{\bot }^2},
\end{equation}
where ${\bf x}_{\bot }\equiv (x_1,x_2,0)$, and ${\bf \hat x}_i$ denotes the
basis vectors in the Cartesian coordinate frame. The constants $q\equiv
-(e_1g_2-e_2g_1)/\hbar c$ and $e^2\equiv -e_1e_2-g_1g_2$ in Eqs. (\ref{2.3})
and (\ref{2.4}) are combinations of the electric and magnetic charges of the
two particles, and $r\equiv \sqrt{x_1^2+x_2^2+x_3^2}$ is the radial
distance, as usual. The hydrogen atom is a special case of the dionium atom
with $e_1=-e_2=e$ and $q=0$. 
An electron around a pure magnetic monopole has
$e_1=e, g_2=g, e_2=g_1=0$.
In the vector potential we have taken the
gauge freedom ${\bf A}\rightarrow {\bf A(x})+\nabla \Lambda ({\bf x})$ to
enforce the transverse gauge $\nabla \cdot {\bf A(x})=0.$ In addition, we have
taken advantage of the extra monopole gauge invariance \cite{2} which allows
us to choose the shape of the Dirac String that imports the magnetic flux to
the monopoles. The field ${\bf A(x})$ in Eq. (\ref{2.4}) has two strings of
equal strength importing the flux, one along the positive $x_3$-axis
from plus infinity to the origin, the other along the negative $x_3$-axis
from minus infinity to the origin. As a
consequence of monopole gauge invariance, the parameter $q$ has to be an
integer or a half-integer number \cite{2}, a condition referred to as Dirac's charge
quantization.

Before time-slicing the path integral, we have to regularize it via a
so-called $f$-transformation \cite{4,7}, which exchanges the path parameter $%
\tau $ by a new one $s$: 
\begin{equation}
\label{2.5}d\tau =dsf_l({\bf x}_n)f_r({\bf x}_{n-1}), 
\end{equation}
where $f_l({\bf x})$ and $f_r({\bf x})$ are invertible functions whose
product is positive. The freedom in choosing $f_{l,r}$ amounts to an
invariance under path-dependent-reparametrizations of the path parameter $%
\tau $ in the fixed-energy amplitude of Eq. (\ref{2.1}). By this transformation,
the (D+1)-dimensional relativistic fixed-energy amplitude for an arbitrary
time-independent potential turns into \cite{4,7} 
\begin{equation}
G({\bf {x}}_b,{\bf {x}}_a;E)\approx \frac{i\hbar }{2Mc}\int_0^\infty
dS\prod_{n=1}^{N+1}\left[ \int d\rho _n\Phi (\rho _n)\right] 
\end{equation}
\begin{equation}
\label{2.6}\times \frac{f_l({\bf x}_a)f_r({\bf x}_b)}{\left( \frac{2\pi
\hbar \epsilon _b^s\rho _bf_l({\bf x}_b)f_r({\bf x}_a)}M\right) ^{D/2}}%
\prod_{n=1}^N\left[ \int_{-\infty }^\infty \frac{d^Dx_n}{\left( \frac{2\pi
\hbar \epsilon _n^s\rho _nf({\bf x}_n)}M\right) ^{D/2}}\right] \exp \left\{
-\frac 1\hbar A^N\right\} 
\end{equation}
with the $s$-sliced action%
$$
A^N=\sum_{n=1}^{N+1}\left[ \frac{M\left( {\bf x}_n-{\bf x}_{n-1}\right) ^2}{%
2\epsilon _n^s\rho _nf_l({\bf x}_n)f_r({\bf x}_{n-1})}-i{\bf A}\cdot ({\bf x}%
_n-{\bf x}_{n-1})\right. 
$$
\begin{equation}
\label{2.7}\left. -\epsilon _n^s\rho _nf_l({\bf x}_n)f_r({\bf x}_{n-1})\frac{%
\left( E-V\right) ^2}{2Mc^2}+\epsilon _n^s\rho _nf_l({\bf x}_n)f_r({\bf x}%
_{n-1})\frac{Mc^2}2\right] . 
\end{equation}
A family of functions which regulates the dionium system is 
\begin{equation}
\label{2.8}f_l({\bf x})=f({\bf {x}})^{1-\lambda },\ \ f_r({\bf x})=f({\ }%
{\bf x})^\lambda , 
\end{equation}
whose product satisfies $f_l({\bf x})f_r({\bf x})=f({\bf x})=r.$ Thus we obtain
the amplitude%
$$
G({\bf {x}}_b,{\bf {x}}_a;E)\approx \frac{i\hbar }{2Mc}\int_0^\infty
dS\prod_{n=1}^{N+1}\left[ \int d\rho _n\Phi (\rho _n)\right] 
$$
\begin{equation}
\label{2.9}\times \frac{r_a^{1-\lambda }r_b^\lambda }{\left( \frac{2\pi
\hbar \epsilon _b^s\rho _br_b^{1-\lambda }r_a^\lambda }M\right) ^{3/2}}%
\prod_{n=2}^{N+1}\left[ \int_{-\infty }^\infty \frac{d^3\triangle x_n}{%
\left( \frac{2\pi \hbar \epsilon _n^s\rho _nr_{n-1}}M\right) ^{3/2}}\right]
\exp \left\{ -\frac 1\hbar A^N\right\} , 
\end{equation}
where the action is%
$$
A^N=\sum_{n=1}^{N+1}\left[ \frac{M\left( {\bf x}_n-{\bf x}_{n-1}\right) ^2}{%
2\epsilon _n^s\rho _nr_n^{1-\lambda }r_{n-1}^\lambda }-i{\bf {A}}_n\cdot (%
{\bf {x}}_n-{\bf {x}}_{n-1})\right. 
$$
\begin{equation}
\label{2.10}\left. -\epsilon _n^s\rho _nr_n\left( r_{n-1}/r_n\right)
^\lambda \frac{(E-V)^2}{2Mc^2}+\epsilon _n^s\rho _nr_n\left(
r_{n-1}/r_n\right) ^\lambda \frac{Mc^2}2\right] . 
\end{equation}
In order to use the Kustaanheimo-Stiefel (KS) transformation (e.g. \cite{4}%
), we now incorporate the dummy fourth dimension into the action by
replacing ${\bf x }$ in the kinetic term by the four-vector $\vec x$
and extending the kinetic action to 
\begin{equation}
\label{2.11}A_{{\rm kin}}^N=\sum_{n=1}^{N+1}\frac M2\frac{\left( \vec
x_n-\vec x_{n-1}\right) }{\epsilon _n^s\rho _nr_n^{1-\lambda
}r_{n-1}^\lambda }.
\end{equation}
This is achieved by satisfaction the 
following trivial identity 
\begin{equation}
\label{2.12}\prod_{n=1}^{N+1}\left[ \int \frac{d(\triangle x^4)_n}{\left(
2\pi \hbar \epsilon _n^s\rho _nr_n^{1-\lambda }r_{n-1}^\lambda /M\right)
^{1/2}}\right] \exp \left\{ -\frac 1\hbar \sum_{n=1}^{N+1}\frac M2\frac{%
(\triangle x_n^4)^2}{\epsilon _n^s\rho _nr_n^{1-\lambda }r_{n-1}^\lambda }%
\right\} =1. 
\end{equation}
Hence the fixed-energy amplitude of the dionium system in three dimensions can
be rewritten as the four-dimensional path integral%
$$
G({\bf {x}}_b,{\bf {x}}_a;E)\approx \frac{i\hbar }{2Mc}\int_0^\infty
dS\prod_{n=1}^{N+1}\left[ \int d\rho _n\Phi (\rho _n)\right] 
$$
\begin{equation}
\label{2.13}\times \int dx_a^4\frac{r_a^{1-\lambda }r_b^\lambda }{\left( 
\frac{2\pi \hbar \epsilon _b^s\rho _br_b^{1-\lambda }r_a^\lambda }M\right) ^2%
}\prod_{n=2}^{N+1}\left[ \int_{-\infty }^\infty \frac{d^4\triangle x_n}{%
\left( \frac{2\pi \hbar \epsilon _n^s\rho _nr_{n-1}}M\right) ^2}\right] \exp
\left\{ -\frac 1\hbar A^N\right\} , 
\end{equation}
where $A^N$ is the action of 
Eq. (\ref{2.10}) in which the three-vectors ${\bf {%
x}}_n$ are replaced by the four-vectors $\vec x_n$. With the help of the
following approximation%
$$
\frac{r_a^{1-\lambda }r_b^\lambda }{\left( \frac{2\pi \hbar \epsilon
_b^s\rho _br_b^{1-\lambda }r_a^\lambda }M\right) ^2}\prod_{n=2}^{N+1}\left[
\int_{-\infty }^\infty \frac{d^4\triangle x_n}{\left( \frac{2\pi \hbar
\epsilon _n^s\rho _nr_{n-1}}M\right) ^2}\right] 
$$
\begin{equation}
\label{2.14}\approx \frac 1{r_a}\frac 1{\left( \frac{2\pi \hbar \epsilon
_b^s\rho _b}M\right) ^2}\prod_{n=2}^{N+1}\left[ \int_{-\infty }^\infty \frac{%
d^4\triangle x_n}{\left( \frac{2\pi \hbar \epsilon _n^s\rho _nr_n}M\right) ^2%
}\right] \exp \left\{ 3\lambda \sum_{n=1}^{N+1}\log \frac{r_n}{r_{n-1}}%
\right\} , 
\end{equation}
where the equality $(r_b/r_a)^{3\lambda
-2}=\prod_1^{N+1}(r_n/r_{n-1})^{3\lambda -2}$ has been used, we obtain%
$$
G({\bf x}_b,{\bf x}_a;E)\approx \frac{i\hbar }{2Mc}\int_0^\infty
dS\prod_{n=1}^{N+1}\left[ \int d\rho _n\Phi (\rho _n)\right] 
$$
\begin{equation}
\label{2.15}\times \int \frac{dx_a^4}{r_a}\frac 1{\left( \frac{2\pi \hbar
\epsilon _b^s\rho _b}M\right) ^2}\prod_{n=2}^{N+1}\left[ \int_{-\infty
}^\infty \frac{d^4\triangle x_n}{\left( \frac{2\pi \hbar \epsilon _n^s\rho
_nr_n}M\right) ^2}\right] \exp \left\{ -\frac 1\hbar \sum_{n=1}^{N+1}\left[
A^N-3\lambda \hbar \log \frac{r_n}{r_{n-1}}\right] \right\} . 
\end{equation}
Since the path integral represents the general relativistic resolvent
operator, all results must be independent of the splitting parameter $%
\lambda $ after going to the continuum limit. Choosing a splitting parameter 
$\lambda =0,$ we obtain the continuum limit of the action 
\begin{equation}
\label{2.16}A_E\left[ x,x^{\prime }\right] =\int_0^Sds\left[ \frac{%
Mx^{\prime 2}}{2\rho r}-i{\bf {A}}\cdot {\bf {x}}^{\prime }-\rho r\frac{%
(E-V)^2}{2Mc^2}+\rho r\frac{Mc^2}2\right] . 
\end{equation}
We now solve the dionium system by introducing the KS transformation (e.g. \cite
{4}) 
\begin{equation}
\label{2.17}d\vec x=2A(\vec u)d\vec u. 
\end{equation}
The arrow on top of $x$ indicates that $\vec{x}$ is a four-vector. For
symmetry reasons, we choose the $4\times 4$ 
matrix $A(\vec u)$ as 
\begin{equation}
\label{2.18}A(\vec u)=\left( 
\begin{array}{cccc}
u^3 & u^4 & u^1 & u^2 \\ 
u^4 & -u^3 & -u^2 & u^1 \\ 
u^1 & u^2 & -u^3 & -u^4 \\ 
u^2 & -u^1 & u^4 & -u^3 
\end{array}
\right).
\end{equation}
The transformation of coordinate difference is 
\begin{equation}
\label{2.19}(\triangle {\bf {x}}_n^i)^2=4{\bf {\bar u}}_n^2(\triangle {\bf {u%
}}_n^i)^2, 
\end{equation}
where ${\bf {\bar u}}_n$ $\equiv ({\bf {u}}_n+{\bf {u}}_{n-1})/2$. In the
continuum limit, this amounts to be
\begin{equation}
\label{2.20}d^4x=16r^2d^4u, 
\end{equation}
\begin{equation}
\label{2.21}\vec x^{\prime 2}=4\vec u^2\vec u^{\prime 2}=4r\vec u^{\prime
2}. 
\end{equation}
By employing the basis tetrad notation $e_{\;\;\mu }^i(\vec u),$ Eq. (\ref
{2.17}) has the form $dx^i=e_{\;\;\mu }^i(\vec u)$ $du^\mu $, 
and it is given by 
\begin{equation}
\label{2.22}e_{\;\;\mu }^i(\vec u)\ =\frac{\partial x^i}{\partial u^\mu }%
(\vec u)=2A_{\;\ \mu }^i(\vec u)\;,\quad i=1,2,3,4. 
\end{equation}
Under the KS transformation, the magnetic interaction turns into 
$$
{\bf {A}}_n\cdot ({\bf {x}}_n-{\bf {x}}_{n-1})=\hbar q\frac{\left[
(x_1)_n(\triangle x_2)_n-(x_2)_n(\triangle x_1)_n\right] (x_3)_n}{r_n{\bf (x}%
_{\bot })_n^2} 
$$
$$
=-\frac{\hbar q}{r_n}
\left[ \vbox to 24pt{} 
\frac{u_n^1\triangle
u_n^2-u_n^2\triangle u_n^1}{\left( u_n^1\right) ^2+\left( u_n^1\right) ^2}+ 
\frac{u_n^4\triangle u_n^3-u_n^3\triangle u_n^4}{\left( u_n^3\right)
^2+\left( u_n^4\right) ^2}
\vbox to 24pt{} \right] 
$$
\begin{equation}
\label{2.23}
\times
\left[ \vbox to 24pt{} 
 \left( u_n^1\right) ^2+\left(
u_n^1\right) ^2-\left( u_n^3\right) ^2-\left( u_n^4\right) ^2
\vbox to 24pt{} \right] . 
\end{equation}
We obtain a path integral equivalent to Eq. (\ref{2.13}) 
\begin{equation}
\label{2.24}G({\bf {x}}_b,{\bf {x}}_a;E)=\frac{i\hbar }{2Mc}\int_0^\infty
dS\ e^{SEe^2/\hbar Mc^2}G(\vec u_b,\vec u_a;S), 
\end{equation}
where $G(\vec u_b,\vec u_a;S)$ denotes the s-sliced amplitude 
\begin{equation}
\label{2.25}\prod_{n=1}^{N+1}\left[ \int d\rho _n\Phi (\rho _n)\right] \frac
1{16}\int \frac{dx_a^4}{r_a}\frac 1{\left( \frac{2\pi \hbar \epsilon
_b^s\rho _b}m\right) ^2}\prod_{n=1}^N\left[ \int_{-\infty }^\infty \frac{%
d^4u_n}{\left( \frac{2\pi \hbar \epsilon _n^s\rho _n}m\right) ^2}\right]
\exp \left\{ -\frac 1\hbar A^N\right\} 
\end{equation}
with the action 
\begin{equation}
\label{2.26}A^N=\sum_{n=1}^{N+1}\left\{ \frac{m(\triangle \vec u_n)^2}{%
2\epsilon _n^s\rho _n}-i(\vec A_n\cdot \triangle \vec u_n)+\epsilon
_n^s\rho _n\frac{m\omega ^2\vec u_n^2}2-\epsilon _n^s\rho _n\frac{\hbar
^24\alpha ^2}{2m\vec u_n^2}\right\} . 
\end{equation}
Here 
\begin{equation}
\label{2.27}m=4M,\quad \omega ^2=\frac{M^2c^4-E^2}{4M^2c^2}, 
\end{equation}
and $\vec A_n\cdot \triangle \vec u_n$ is given by Eq. (\ref{2.23}). After
taking the continuum limit, this leads to the Duru-Kleinert transformed
action 
\begin{equation}
\label{2.28}A=\int_0^Sds\left[ \frac{m\vec u^{\prime 2}}{2\rho (s)}-i(\vec
A\cdot \vec u^{\prime })+\rho (s)\frac{m\omega ^2\vec u^2}2-\rho (s)\frac{%
4\hbar ^2\alpha ^2}{2m\vec u^2}\right] . 
\end{equation}

Let us now analyze the
effect coming from the magnetic interaction upon the Coulomb system. We first
express $(u^1,u^2,u^3,u^4)$ in terms of three-dimensional spherical
coordinate with an auxiliary angle $\gamma $: 
\begin{equation}
\label{2.32}\left. 
\begin{array}{l}
u^1= 
\sqrt{r}\cos (\theta /2)\cos \left[ (\varphi +\gamma )/2\right] \\ u^2= 
\sqrt{r}\cos (\theta /2)\sin \left[ (\varphi +\gamma )/2\right] \\ u^3= 
\sqrt{r}\sin (\theta /2)\cos \left[ (\varphi -\gamma )/2\right] \\ u^4= 
\sqrt{r}\sin (\theta /2)\sin \left[ (\varphi -\gamma )/2\right] 
\end{array}
\right\} \quad \left( 
\begin{array}{l}
0\leq \theta \leq \pi \\ 
0\leq \varphi \leq 2\pi \\ 
0\leq \gamma \leq 4\pi 
\end{array}
\right) . 
\end{equation}
This gives us the equivalent form of Eq. (\ref{2.25}) 
\begin{equation}
\label{2.33}\prod_{n=1}^{N+1}\left[ \int d\rho _n\Phi (\rho _n)\right] \frac
1{16}\int_0^{4\pi }d\gamma _a\frac 1{\left( \frac{2\pi \hbar \epsilon
_b^s\rho _b}m\right) ^2}\prod_{n=1}^N\left[ \int_{-\infty }^\infty \frac{%
d^4u_n}{\left( \frac{2\pi \hbar \epsilon _n^s\rho _n}m\right) ^2}\right]
\exp \left\{ -\frac 1\hbar A^N\right\} 
\end{equation}
with the continuum action 
$$
A=\int_0^Sds\left\{ \frac m{2\rho (s)}\left[ u^{\prime 2}+\frac{u^2}4\left(
\theta ^{\prime 2}+\varphi ^{\prime 2}+\gamma ^{\prime 2}+2\varphi ^{\prime
}\left( \gamma ^{\prime }-\rho \frac{4\hbar qi}{mu^2}\right) \cos \theta
\right) \right] \right. 
$$
\begin{equation}
\label{2.34}\left. +\rho (s)\frac{m\omega ^2\vec u^2}2-\rho (s)\frac{4\hbar
^2\alpha ^2}{2m\vec u^2}\right\} . 
\end{equation}
The spherical coordinate $(u,\theta ,\varphi )$ and the auxiliary angle $%
\gamma $ can be represented by canonical momentums as follows: 
\begin{equation}
\label{2.35} 
\left \{
\begin{array}{l}
u^{\prime }=\rho 
\frac{p_u}m, \\ \theta ^{\prime }= 
\frac{p_\theta }\xi , \\ \gamma ^{\prime }=\frac 1{\xi \sin {}^2\theta
}\left[ p_\gamma +\eta \xi \cos {}^2\theta -p_\varphi \cos \theta
\right] , \\ 
\varphi ^{\prime }=\frac 1{\xi \sin {}^2\theta }\left[ p_\varphi -\eta
\xi \cos {}\theta -p_\gamma \cos \theta \right] , 
\end{array} 
\right.
\end{equation}
where the variables $\eta \equiv -4\rho \hbar qi/mu^2$ and $\xi \equiv
mu^2/4\rho .$ 
With the help of Eq. (\ref{2.35}), 
we obtain the canonical form of the path integral%
$$
\frac{i\hbar }{2Mc}\int_0^\infty dS\ e^{SEe^2/\hbar
Mc^2}\prod_{n=1}^{N+1}\left[ \int d\rho _n\Phi (\rho _n)\right] \frac
1{16}\int_0^{4\pi }d\gamma _a 
$$
\begin{equation}
\label{2.36}\times \prod_{n=1}^N\left[ \int_{-\infty }^\infty d^4\vec
u_n\right] \prod_{n=1}^{N+1}\left[ \int_{-\infty }^\infty \frac{d^4(p_u)_n}
{2\pi \hbar} \right]
\exp \left\{ -\frac 1\hbar A^N\right\} 
\end{equation}
with the action of continuum version 
\begin{equation}
\label{2.37}A=\int_0^Sds\left\{ -i\left[ p_uu^{\prime }+p_\theta \theta
^{\prime }+p_\varphi \varphi ^{\prime }+(p_\gamma -\hbar q)\gamma ^{\prime
}\right] +H\right\} , 
\end{equation}
where the Hamiltonian is given by 
$$
H=\frac \rho {2m}\left\{ p_u^2+\frac 4{\vec u^2}\left[ p_\theta ^2+\frac
1{\sin {}^2\theta }\left( p_\varphi ^2+p_\gamma ^2-2p_\gamma p_\varphi \cos
\theta \right) \right] \right\} 
$$
\begin{equation}
\label{2.38}+\frac{4\rho }{2m\vec u^2}\left[ -2\hbar q\left( p_\gamma -\hbar
q\right) +\hbar ^2\left( \alpha ^2+q^2\right) \right] +\rho \frac{m\omega
^2\vec u^2}2. 
\end{equation}
This differs from the pure relativistic Coulomb system \cite{6} in two
places:

First, the Hamiltonian has an extra centrifugal barrier proportional to the
charge parameter $q$: 
\begin{equation}
\label{2.39}V(r)=\frac{-2\hbar q \rho \left( p_\gamma -\hbar q\right) }{2Mr}. 
\end{equation}
Second, the action of Eq. (\ref{2.37}) contains an additional term 
\begin{equation}
\label{2.40}\triangle A=-\hbar q\int_0^Sds\gamma ^{\prime }. 
\end{equation}
Fortunately, this is a pure surface term $\triangle A=-\hbar q(\gamma
_b-\gamma _a).$ Due to Eq. (\ref{2.40}), the modification consists of a
simple extra phase factor in the integral over $\gamma _a$ so that 
\begin{equation}
\label{2.41}G({\bf {x}}_b,{\bf {x}}_a;E)=\frac{i\hbar }{2Mc}\int_0^\infty
dS\ e^{SEe^2/\hbar Mc^2}\frac 1{16}\int_0^{4\pi }d\gamma _ae^{-iq(\gamma
_b-\gamma _a)}G(\vec u_b,\vec u_a;S). 
\end{equation}
Since the integral over $\gamma _a$ forces the momentum $p_r$ in the
canonical action (\ref{2.37}) to take the value $\hbar q.$ This eliminates
the term proportional to $p_\gamma -\hbar q$ in Eq. (\ref{2.38}), therefore
the Green's function in $u$-space has the form 
\begin{equation}
\label{2.42}G(\vec u_b,\vec u_a;S)\approx \prod_{n=1}^{N+1}\left[ \int d\rho
_n\Phi (\rho _n)\right] \frac 1{\left( \frac{2\pi \hbar \epsilon _b^s\rho _b}%
m\right) ^2}\prod_{n=1}^N\left[ \int_{-\infty }^\infty \frac{d^4u_n}{\left( 
\frac{2\pi \hbar \epsilon _n^s\rho _n}m\right) ^2}\right] \exp \left\{
-\frac 1\hbar A^N\right\} 
\end{equation}
with the action 
\begin{equation}
\label{2.43}A^N=\sum_{n=1}^{N+1}\left\{ \frac{m(\triangle \vec u_n)^2}{%
2\epsilon _n^s\rho _n}+\epsilon _n^s\rho _n\frac{m\omega ^2\vec u_n^2}%
2-\epsilon _n^s\rho _n\frac{\hbar ^24(\alpha ^2+q^2)}{2m\vec u_n^2}\right\}
. 
\end{equation}
We now choose the gauge $\rho (s)=1$ in Eq. (\ref{2.42}). We arrive at
\begin{equation}
\label{2.44}G(\vec u_b,\vec u_a;S)\approx \frac 1{\left( \frac{2\pi \hbar
\epsilon _b^s}m\right) ^2}\prod_{n=1}^N\left[ \int_{-\infty }^\infty \frac{%
d^4u_n}{\left( \frac{2\pi \hbar \epsilon _n^s}m\right) ^2}\right] \exp
\left\{ -\frac 1\hbar A^N\right\} , 
\end{equation}
where the action 
\begin{equation}
\label{2.55}A^N=\sum_{n=1}^{N+1}\left\{ \frac{m(\triangle \vec u_n)^2}{%
2\epsilon _n^s}+\epsilon _n^s\frac{m\omega ^2\vec u_n^2}2-\epsilon _n^s\frac{%
\hbar ^24(\alpha ^2+q^2)}{2m\vec u_n^2}\right\} . 
\end{equation}
It describes a particle
with mass $m=4M$ moving as a function of 
``pseudotime'' $s$ in a 4-dimensional harmonic oscillator potential of
frequency 
\begin{equation}
\label{2.29}\omega ^2=\frac{M^2c^4-E^2}{4M^2c^2}. 
\end{equation}
The oscillator possesses an additional attractive potential $-4\hbar
^2(\alpha ^2+q^2)/2m\vec u^2$ which is conveniently parametrized in the form of a
centrifugal barrier 
\begin{equation}
\label{2.30}V_{{\rm extra}}=\hbar ^2\frac{l_{{\rm extra}}^2}{2m\vec u^2}, 
\end{equation}
whose squared angular momentum has the negative value $l_{{\rm extra}%
}^2\equiv -4(\alpha ^2+q^2)$, where $\alpha $ denotes the fine-structure constant $%
\alpha \equiv e^2/\hbar c$.

There are no $\lambda $-slicing corrections. This is ensured by the affine
connection of KS transformation satisfying 
\begin{equation}
\label{2.31}\Gamma _\mu ^{\;\;\;\mu \lambda }=g^{\mu \nu }e_i^{\;\;\lambda
}\partial _\mu e_{\;\;\nu }^i=0 
\end{equation}
and the transverse gauge $\partial _iA^i=0$ \cite{4,7}. 
The path integral Eq. (\ref{2.44}) 
can be performed and is given by \cite{4}%
$$
G(\vec u_b,\vec u_a;S)=\frac 1{u_bu_a}\sum_{l=0}^\infty G(u_b,u_a;S,l) 
$$
\begin{equation}
\label{2.56}\times \frac{l+1}{2\pi ^2}%
\sum_{k_1,k_2=-l/2}^{l/2}d_{k_1,k_2}^{l/2}(\theta
_b)d_{k_1,k_2}^{l/2}(\theta _a)e^{ik_1(\varphi _b-\varphi _a)+ik_2(\gamma
_b-\gamma _a)}
\end{equation}
with the radial amplitude 
\begin{equation}
\label{2.57}G(u_b,u_a;S,l)=\frac{m\omega }{\hbar \sinh \omega s}e^{-\frac{%
m\omega }{2\hbar }\left( u_b^2+u_a^2\right) \coth \omega s}I_{\sqrt{%
(l+1)^2-4(\alpha ^2+q^2)}}\left( \frac m\hbar \frac{\omega u_bu_a}{\sinh
\omega s}\right) , 
\end{equation}
where $d_{k_1,k_2}^{l/2}(\theta)e^{ik_1 \varphi +ik_2\gamma}$
are the representation functions of the rotation group (e.g.\cite{4}).
The integral $\int_0^{4\pi }d\gamma _ae^{-iq(\gamma _b-\gamma _a)}$ in Eq. (%
\ref{2.41}) now can be easily done. We arrive at the fixed-energy amplitude
of the relativistic dionium atom, labeled by the subscript $D,$%
\begin{equation}
\label{2.58}G({\bf {x}}_b,{\bf {x}}_a;E)=\frac
1{\sqrt{r_br_a}}\sum_{l_D}G(r_b,r_a;E_D,l_D)\sum_{k=-l_D}^{l_D}Y_{l_D,k,q}(\theta
_b,\varphi _b)Y_{l_D,k,q}^{\star }(\theta _a,\varphi _a), 
\end{equation}
where $Y_{l_D,k,q}(\theta _b,\varphi _b)$ are the so-called monopole
harmonics 
\begin{equation}
\label{2.59}Y_{l_D,k,q}(\theta ,\varphi )=\sqrt{\frac{l+1}{4\pi }}%
e^{ik\varphi }d_{k,q}^{l_D}(\theta ), 
\end{equation}
and $l_D$ is defined as $l/2$. The radial amplitude for the dionium is%
$$
G(r_b,r_a;E_D,l_D)=\frac{i\hbar }{2Mc}\frac 12\int_0^\infty dS\
e^{SE_De^2/\hbar Mc^2} 
$$
\begin{equation}
\label{2.60}\times \frac{m\omega }{\hbar \sinh \omega s}e^{-\frac{m\omega }{%
2\hbar }\left( r_b+r_a\right) \coth \omega s}I_{\sqrt{(2l_D+1)^2-4(\alpha
^2+q^2)}}\left( \frac{m\omega \sqrt{r_br_a}}{\hbar \sinh \omega s}\right) . 
\end{equation}
This integral can be calculated by employing the formula%
$$
\int_0^\infty dy\frac{e^{2\nu y}}{\sinh y}\exp \left[ -\frac t2\left( \zeta
_a+\zeta _b\right) \coth y\right] I_\mu \left( \frac{t\sqrt{\zeta _b\zeta _a}
}{\sinh y}\right) 
$$
\begin{equation}
\label{2.61}=\frac{\Gamma \left( \left( 1+\mu \right) /2-\nu \right) }{t 
\sqrt{\zeta _b\zeta _a}\Gamma \left( \mu +1\right) }W_{\nu ,\mu /2}\left(
t\zeta _b\right) M_{\nu ,\mu /2}\left( t\zeta _b\right) , 
\end{equation}
with the range of validity%
$$
\begin{array}{l}
\zeta _b>\zeta _a>0, \\ 
{\rm Re}[(1+\mu )/2-\nu ]>0, \\ {\rm Re}(t)>0,\mid \arg t\mid <\pi , 
\end{array}
$$
where $M_{\mu ,\nu }$ and $W_{\mu ,\nu }$ are the Whittaker functions, we
complete the integration of Eq. (\ref{2.60}), and find the amplitude for $%
u_b>u_a$ in the closed form, 
\newpage
$$
G(r_b,r_a;E_D,l_D)=\frac{i\hbar }{2Mc}\frac 1{2\omega }\frac{\Gamma \left(
\frac 12+\frac 12\sqrt{(2l_D+1)^2-4(\alpha ^2+q^2)}-E_D\alpha /\sqrt{%
M^2c^4-E_D^2}\right) }{\sqrt{r_br_a}\Gamma \left( \sqrt{(2l_D+1)^2-4(\alpha
^2+q^2)}+1\right) } 
$$
$$
\times W_{E_D\alpha /\sqrt{M^2c^4-E_D^2},\sqrt{(2l_D+1)^2-4(\alpha ^2+q^2)}%
/2}\left( \frac 2{\hbar c}\sqrt{M^2c^4-E_D^2}r_b\right) 
$$
\begin{equation}
\label{2.62}\times M_{E_D\alpha /\sqrt{M^2c^4-E_D^2},\sqrt{(2l_D+1)^2-4(\alpha
^2+q^2)}/2}\left( \frac 2{\hbar c}\sqrt{M^2c^4-E_D^2}r_a\right) . 
\end{equation}
It is easy to check, with $n_r=n-l_D-1$, that the spectra reduced to relativistic
Coulomb case, if we take $q=0, e_1=-e_2=e$ \cite{6}.

The energy spectra can be extracted from the poles. They are determined by 
\begin{equation}
\label{2.63}\frac 12+\frac 12\sqrt{(2l_D+1)^2-4(\alpha ^2+q^2)}-E_D\alpha / 
\sqrt{M^2c^4-E_D^2}=-n_r,\quad n_r=0,1,2,3,\cdots .
\end{equation}
After some mathematical manipulation, we have
\begin{equation}
E_{n_r,l_D,q}=\pm Mc^2\left[ 1+\frac{\alpha ^2}
{\left( \frac 12+\frac 12\sqrt
{\left( 2l_D+1\right) ^2-
4\left( q^2+\alpha ^2\right) }+n_r\right) ^2}\right] ^{-1/2}.
\end{equation}
\\
\centerline{ACKNOWLEDGMENTS}
\\
The author is grateful to Doctors  
M. C. Chang and X. Xu who 
critically read the entire manuscript.

\end{document}